\documentclass[aps,pra,
twocolumn,
showpacs]{revtex4}

\usepackage{amsmath}
\usepackage{epstopdf}
\usepackage{color}

\usepackage{epsfig}
\usepackage{graphicx}

\usepackage{amssymb}
\usepackage{amsfonts}
\usepackage{dcolumn}

\newcommand{\bM}{{\bf M}}

\newcommand{\sn}{\mathrm{sn}}
\newcommand{\dn}{\mathrm{dn}}

\newcommand{\tq}{\tilde{q}}

\newcommand{\PT}{\mathcal{PT}}

\begin{document}


\title{Nonlinear  currents in a ring-shaped waveguide with balanced  gain and dissipation}


\author{Dmitry A. Zezyulin and Vladimir V. Konotop}

\affiliation{Centro de F\'isica Te\'orica e Computacional and Departamento de F\'isica, Faculdade de Ci\^encias, Universidade de Lisboa, Campo Grande 2, Edif\'icio C8, Lisboa 1749-016, Portugal
}

\begin{abstract}
We describe linear and nonlinear modes in a ring-shaped waveguide with localized gain and  dissipation modeled by two  Dirac $\delta$ functions located symmetrically.
The strengths of the gain and dissipation are  equal, i.e., the system obeys the parity-time symmetry. This configuration   describes atomic Bose-Einstein condensates with local loading and local elimination of atoms, polaritonic condensates,  or  optical ring resonators with local pump and absorption. We discuss the linear spectrum of such a system and show if the  location of the $\delta$ functions is slightly asymmetric,  then the system can be driven through a series of  exceptional points by the change  of the gain-and-loss coefficient. In the nonlinear case, the system admits solutions with spatially constant and periodic densities which are presented  in the exact analytical form. These solutions are supported by the current directed from the  gain potential towards the absorbing potential. The system also admits currentless solutions.   Stability and bifurcations of nonlinear solutions are also described.
\end{abstract}


\maketitle

\section{Introduction}

Strongly localized dissipative or active potentials are known to produce sometimes counter-intuitive effects in nonlinear systems. These include  localized states of Bose-Einstein condensates (BECs) in  optical lattices due to the dissipative losses at the boundaries~\cite{Oppo1}; excitation and  supporting of bright and dark solitons in an array of BECs on a line~\cite{BKPO} and on a ring~\cite{Oppo2} by localized dissipation; suppression of the decay of BECs~\cite{ZK} by local elimination of atoms confirmed experimentally in~\cite{Ott}; jamming anomaly  when localized dissipation is compensated by localized gain (spatially separated from the dissipation)~\cite{BarZezKon}, to mention but a few.

On the other hand, in the linear physics localized dissipative or active potentials can admit   the so-called spectral singularities  which are  known as the special points of continuum spectra of    non-hermitian potentials, where  the bi-orthonormal bases of the respective Hamiltonians lose their completeness~\cite{Mostafa_review}. In physical terms, the spectral singularities   correspond to  zero-width resonances~\cite{Zero_resonance} at which the system can act as a laser. Defining the scattering matrix of such a potential by  $\bM$ and the wavenumber of the incident radiation by $q$, the spectral singularity $q_0$ is defined by the real zero of the matrix element $M_{22}(q)$, i.e., $M_{22}(q_0)=0$. In a similar way, a real zero of the element  $M_{11}(q)$  is referred to as  a time-reversed spectral singularity. In the latter case,  the system acts as a coherent perfect absorber (CPA)~\cite{CPA}. For a generic complex potential, the spectral singularities and time-reversed spectral singularities do not occur simultaneously.
If however the potential obeys the parity-time ($\PT$) symmetry~\cite{Bender} (see also~\cite{KYZ} for a recent review), then  the mentioned singularities can not only exist simultaneously but also  coincide. In this   case, the spectral singularities are termed self-dual~\cite{Mostafa_self_dual} and physically correspond to the so-called a CPA-laser~\cite{CPA-laser}, i.e., the potential acts as a CPA and as a laser at the same wavelength.

The above-mentioned facts about spectral singularities characterize the linear spectrum of a potential  and, as such, are pertinent to linear systems. The concept can be   generalized to 
systems where the active guiding medium
is characterized by  nonlinearity   with finite support~\cite{Mostafa_nonlinear} (i.e., sharply localized in space). This is, for example, the case of the stationary light scattering by a slab with the Kerr nonlinearity.

In the present study,  we address  yet another situation. We consider a pair of   absorbing and active {\em linear} potentials embedded in a {\em nonlinear} defocusing medium.
We assume that when the active potential is considered  alone (separately from the absorbing one) 
it is characterized by the spectral singularity $q_0$:  $M^+_{22}(q_0)=0$, where $\bM^+$ is the scattering matrix of the active   potential. Additionally, the absorbing potential, when considered separately from the active one, has the  time-reversed spectral singularity at the same wavelength: $M^-_{11}(q_0)=0$,  where $\bM^-$ is the scattering matrix of the absorbing potential.
%
%
%
When the acting and absorbing potentials act simultaneously on the infinite line, their combined action on an incident    wave    is characterized by the transfer matrix $\bM=\bM^+\bM^-$ (or $\bM=\bM^-\bM^+$, depending on the location of the potentials with respect to the direction of the incidence). The combined transfer matrix ${\bf M}$ does not manifest properties of either laser or a CPA at $q_0$ because, generically, $M_{22}(q_0)\neq 0 $ and $M_{11}(q_0)\neq 0 $. Additionally, in a finite system, like the ring-shaped cavity we study below, the linear spectrum is discrete. Therefore in our system $q_0$ is not a (time-reversed) spectral singularity in the strict mathematical sense.   However, as we will show, the system admits  linear  and nonlinear  solutions in the form of  plane waves which are emitted and absorbed at the wavelength prescribed by the spectral singularities $q_0$.  Bearing these  solutions in mind, we say that our system has   a \emph{pseudo}-singularity $q_0$. The study of solutions supported by such pseudo-singularities is the main goal of this work.

The paper is organized as follows. After formulating the statement of the problem in Sec.~\ref{sec:model}, in Sec.~\ref{sec:linear} we address its  linear properties and show that the system can be manipulated to pass through a series of exceptional points.
In Sec.~\ref{sec:const_ampl} we show that the system   admits    a nonlinear constant-amplitude solution emitted and absorbed on the wavelength defined by the values of the spectral singularities of active and absorbing potentials.    More general, periodic-density solutions are discussed in Sec.~\ref{sec:periodic}. Outcomes are briefly discussed in Conclusion.

%

\section{The model}
\label{sec:model}

\begin{figure}
 		\includegraphics[width=0.95\textwidth]{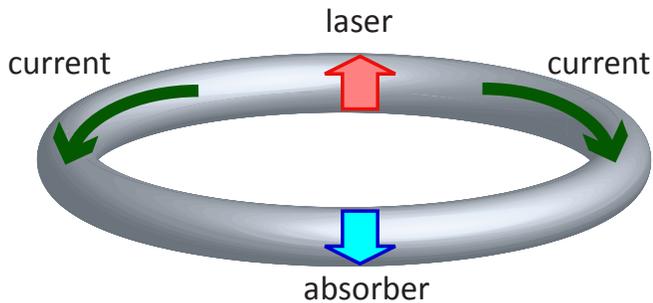}
 	\vspace{-5cm}
 	\caption{(Color online) Schematic representation of lasing (upward pointing arrow) and absorbing (downward pointing arrow) localized potentials placed in a quasi-one-dimensional ring-shaped waveguide filled with a defocusing nonlinear medium.}
 	\label{fig:one}
 \end{figure}

In order to make a spectral singularity to work as a laser, i.e.,   to emit waves without any absorption (or absorb without any emission, in the case of time-reversed singularity), this emission (absorption) should occur from both sides of the respective localized potential~\cite{CPA-laser}. This means that two localized potentials, one emitting  and another absorbing waves, being placed on the infinite line would require ingoing and outgoing energy flows from the infinities. Similar problems  of stabilizing absorption in a nonlinear media by inflows from the infinity have been considered previously~\cite{ZK}. Here we are interested in a system which is  self-sustained, i.e., does not require application of external currents (notice however that the system remains open since the gain potential usually requires  external pumping). This can be achieved by placing lasing and absorbing potentials on a guiding ring as schematically illustrated in Fig.~\ref{fig:one}. If lasing and absorption occur at the same wavelength, the two (left and right) flows generated by the ``laser'' (active potential) propagate towards the absorber  (dissipative potential) which annihilates the flows.

The described setting can be implemented in several physical systems. This can be a BEC  in a toroidal trap~\cite{toroid}, with atoms loaded at one side by an atomic laser~\cite{atom_laser} and eliminated from the opposite side, say by electronic beams~\cite{Ott}. This can be a ring-shaped exciton-polariton condensate~\cite{ring_shaped_polarit} nonresonantly pumped from the one side and having enhanced losses at the other side (the statement similar to the experiment~\cite{polariton}). Yet another system described by our model is a ring-shaped optical waveguide with active and passive elements, similarly to the design of the recent experiments on the microring lasers~\cite{lasers}.

Considering an atomic BEC in a sufficiently narrow toroidal trap (see \cite{BludKon} for the discussion of physical applicability of this approximation), the mathematical formulation of the model is reduced to the Gross-Pitaevskii (GP) equation with complex potentials having the Dirac delta function shapes:
\begin{eqnarray}
\label{GPE}
i\Psi_{t}=-\Psi_{\theta\theta}+|\Psi|^2\Psi+2iq_0\delta(\theta\textcolor{black}{-\nu})\Psi - 2iq_0\delta(\theta+\pi)\Psi.
\end{eqnarray}
Here $\Psi$ is the macroscopic wavefunction, $\theta\in[-\pi,\pi)$ is the angular coordinate, $q_0>0$ is a constant characterizing (equal) strengths of the gain at $\theta=\textcolor{black}{\nu\in [0, \pi)}$ and loss at $\theta=\pi$, and the repulsive (defocusing) nonlinearity is scaled out to one. \textcolor{black}{The case $\nu=0$ corresponds to the situation when the two $\delta$ functions are placed     diametrically opposite each other   (this is the situation illustrated in Fig.~\ref{fig:one})}

Solutions of the nonlinear problem (\ref{GPE})  can be searched in the intervals $\theta\in[-\pi,\textcolor{black}{\nu})$ and $\theta\in(\textcolor{black}{\nu},\pi)$   separately: we denote them by $\Psi_1(\theta,t)$ and $\Psi_2(\theta,t)$, respectively. These solutions are  connected by the matching 
conditions:
\begin{subequations}
 \label{boundary}
\begin{eqnarray}
\label{boundary_1}
&&\begin{array}{l}
\Psi_1(\textcolor{black}{\nu},t)=\Psi_2(\textcolor{black}{\nu},t),
\\
\Psi_{2,\theta}(\textcolor{black}{\nu},t)-\Psi_{1,\theta}(\textcolor{black}{\nu},t)=2iq_0 \Psi_{1,2}(\textcolor{black}{\nu},t),
\end{array}
\\
\label{boundary_2}
&&\begin{array}{l}
\Psi_1(-\pi,t)=\Psi_2(\pi,t),
\\
\Psi_{1,\theta}(-\pi,t)-\Psi_{2,\theta}(\pi,t)=-2iq_0 \Psi_{1,2}(\mp \pi,t).
\end{array}
\end{eqnarray}
\end{subequations}

A model similar to (\ref{GPE})  with a $\PT$-symmetric double-well potential mimicked by $\delta$ functions with complex amplitudes, but considered on the infinite  line, was addressed in~\cite{Cart,BZ}, from the point of view of the existence of the nonlinear modes, and in~\cite{Cartarius}  {spectral singularities of such double-delta potentials were computed numerically}. Besides the different geometry, the difference between our model (\ref{GPE}) in that considered in \cite{Cartarius,Cart,BZ} is that in our case the amplitudes of $\delta$ functions are purely imaginary, i.e., our delta functions describe  only gain and  losses, without any conservative potential wells of barriers.

We recall that purely  imaginary $\delta$ potentials $2iq_0\delta(\theta\textcolor{black}{-\nu})$ and $-2iq_0\delta(\theta+\pi)$ considered separately  admit  spectral singularity and time-reversed spectral singularity, respectively,  at the same wavevector $q_0$~\cite{Mostafa_delta}. In the model (\ref{GPE}), these singularities   are responsible for the laser at $\theta=\textcolor{black}{\nu}$ and  for CPA at $\theta=-\pi$. 
Thus, in our setting $q_0$ is the pseudo-singularity. 

\section{Linear spectrum and exceptional points}
\label{sec:linear}

{\color{black} Before we proceed with the analysis of the nonlinear modes  supported by Eq.~(\ref{GPE}), it is instructive to understand the properties of its linear counterpart. On the one hand, this appears important since  the effect of nonlinearity is expected to vanish for small-amplitude nonlinear modes, and therefore the linear case can give some useful {\it a priori} information about the small-amplitude limit. On the other hand,  the concept of pseudo-singularity originates in  the linear theory, so the linear case must be understood before we consider the possible nonlinear generalizations. Additionally,  the analysis of the ring geometry with balanced   gain and loss was not reported so far, so it appears timely to inspect this problem. }

	
Making the substitution $\Psi(\theta, t) = \psi(\theta) e^{-i \tq^2 t}$ and omitting the nonlinear term in (\ref{GPE}), we arrive at the linear eigenvalue problem (hereafter tilde indicates the linear eigenvalues):
 \begin{eqnarray}
 \label{GPE_linear}
 \tq^2\psi=-\psi_{\theta\theta}+2iq_0\delta(\theta -\nu)\psi - 2iq_0\delta(\theta+\pi)\psi.
 \end{eqnarray}
The linear chemical potential $\tilde{\mu}$ amounts to   $\tilde{\mu}=\tq^2$, and thus real or purely imaginary $\tq$ correspond to the  stable linear waves. Denoting the linear eigenfunctions $\psi_1(\theta;\tq)$ and $\psi_2(\theta,\tq)$, in the intervals $\theta\in(\nu,\pi)$ and $\theta\in(-\pi,\nu)$,
the spectrum of (\ref{GPE_linear}) 
is found by the substitution
 \begin{eqnarray}
  \psi_j=c_{j1}e^{i\tq\theta}+c_{j2}e^{-i\tq\theta}.
 \end{eqnarray}
Using the matching conditions    (\ref{boundary}), we obtain the
following  transcendental equation:
 \begin{eqnarray}
 \label{dispersion}
 \tq^2-\tq^2\cos(2\pi \tq)+q_0^2\cos(2\pi \tq)-q_0^2\cos(2\tq \nu)=0.
 \end{eqnarray}

\begin{figure}
   	\includegraphics[width=0.9\columnwidth]{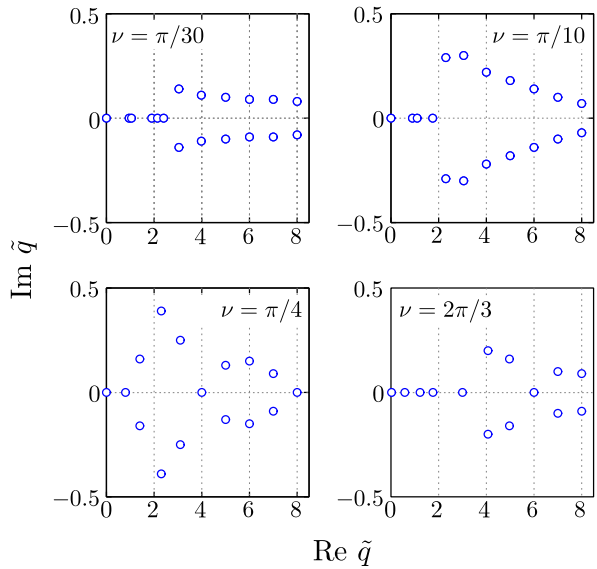}
   	\caption{(Color online) \textcolor{black}{Several smallest amplitude eigenvalues $\tilde{q}$ for fixed value of the gain-and-loss strength $q_0=5/2$ and four increasing values of the angle $\nu$. Only eigenvalues with $\textrm{Re}\,\tq\geq0$ are shown.}
   }
   	\label{fig:spectra}
   \end{figure}

It follows from (\ref{dispersion}) that at $\nu=0$ the spectrum is purely real for any $q_0$.  If $q_0$ is not an integer, then the spectrum for $\nu=0$ contains  one { simple} eigenvalue $\tq_0 = q_0$ and a set of {  double} eigenvalues $\tq_n = n$, where $n$ runs through all positive integers.  For small $\nu\ll 1 $ solutions of (\ref{dispersion}) can be found by the expansion in terms of $\nu$:
 \begin{eqnarray}
 \label{spectrum_0}
&& \tq_0(\nu)=q_0-\nu^2 \frac{q_0^2}{1-\cos(2\pi q_0)}+ \mathcal{O}(\nu^4), \quad q_0\notin \mathbb{N}
 \\
  \label{spectrum_n}
 && \tq_n(\nu)=n\pm \nu^2\frac{nq_0}{\pi\sqrt{q_0^2-n^2}} +\mathcal{O}(\nu^2), \quad n\in\mathbb{N}^+
 \end{eqnarray}
 Thus under small perturbation in a form of a shift of one of the $\delta$-potentials (by the angle $\nu$ in our case), the simple eigenvalue is shifted along the real axis, while each double eigenvalue    splits into two simple ones. If  $n<q_0$, then the split  eigenvalues  remain real, whereas for $n>q_0$ the eigenvalues acquire nonzero imaginary parts.
 \textcolor{black}{For arbitrary values of $\nu$, the spectrum of eigenvalues $\tilde{q}$ can be obtained numerically by solving  Eq.~(\ref{dispersion}) directly. Several obtained eigenvalue portraits for the fixed value of the gain-and-loss strength $q_0=5/2$ are presented in Fig.~\ref{fig:spectra} As $\nu$ increases, the eigenvalues  undergo various transitions. Some of them  collide   pairwise at the real axis and acquire nonzero imaginary parts. At the same time, for some values of $\nu$ some previously    complex eigenvalues can become real: for example, for $\nu=\pi/4$  one can find real eigenvalues at $\tq=4$ and $\tq=8$. However, the spectrum is never entirely real for any nonzero $\nu$.}

\begin{figure}
   	\includegraphics[width=\columnwidth]{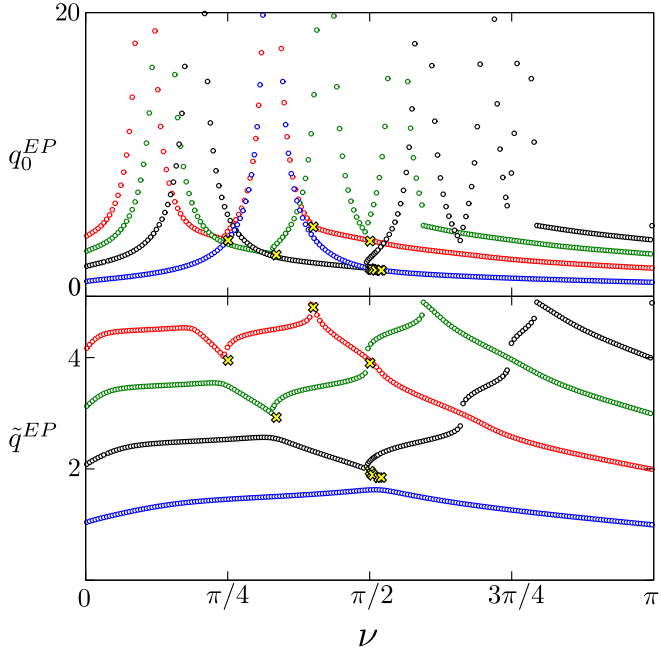}
   	\caption{(Color online) \textcolor{black}{Pseudo-singularities  $q_0^{EP}$ generating the exceptional points  (upper panel) and the double eigenvalues $\tq^{EP}$ corresponding to these exceptional points (lower panel)
     {\it vs} the shift $\nu$ of the lasing $\delta$-potential. The most of the shown double eigenvalues and the exceptional points correspond to the situation   for which  the colliding eigenvalues are complex for $q_0<q_0^{EP}$ but become real for $q_0>q_0^{EP}$. Positions of double eigenvalues and exceptional points for which  the colliding eigenvalues are real for $q_0<q_0^{EP}$ but become complex-conjugates for $q_0>q_0^{EP}$ are labeled with the crosses.}}
   	\label{fig:except}
   \end{figure}

Turning to the effect of changing of the of gain-and-loss strength, expansions  (\ref{spectrum_0})--(\ref{spectrum_n}) suggest an interesting observation:
%
the increase of gain and loss (i.e., of the   pseudo-singularity $q_0$) can partially improve stability   shifting the unstable modes to higher wavelengths. For general  (not necessarily small) value of $\nu$, one can change (increase or decrease) the pseudo-singularity $q_0$ to make two    eigenvalues   coalesce at the exceptional point and exit to the complex plane giving raise to new unstable modes (or to come from the complex plane to the real axis transforming in this way unstable modes to stable). The location of the colliding eigenvalues $\tq$   can be obtained as  zeros of the dispersion relation (\ref{dispersion}) which  are simultaneously zeros of the derivative of (\ref{dispersion}) with respect to $\tq$. The compatibility condition for those two equations readily gives the implicit relation between $\tq$ and $\nu$:
  \begin{eqnarray}
  \label{excet_eq}
  \sin(\pi \tq)[q\nu\sin(2 \tq\nu)+\cos(2 \tq\nu){-}\cos(2 \tq\pi)]
  \nonumber\\
  =\tq\pi\cos(\pi \tq) [1-\cos(2 \tq\nu)]
  \end{eqnarray}
whose solutions yield the positions of the colliding eigenvalues  $\tq = \tq^{EP}$   for the given shift $\nu$ (the upper index $EP$ stands for the exceptional point). The exceptional points correspond  to the  pseudo-singularities $q_0=q_0^{EP}$   given by the relation
\begin{eqnarray}
  \label{k0_except}
  q_0^{EP}=\frac{\tq^{EP}\sin(\pi \tq^{EP})}{\left\{\sin[\tq^{EP}(\pi+\nu)]\sin[\tq^{EP}(\pi-\nu)]\right\}^{1/2}}.
 \end{eqnarray}
In order to find the exceptional points, we solve numerically Eq.~(\ref{excet_eq}) for different values of $\nu$ \textcolor{black}{increasing  from $\nu=0$ to $\nu=\pi$ with   step $\Delta\nu=\pi/200$} and  obtain the location of colliding eigenvalues $\tq=\tq^{EP}$, see the \textcolor{black}{lower} panel in Fig.~\ref{fig:except} (the computation has been limited by the search of the  colliding eigenvalues   inside the finite interval   $\tq \in [0.1,5]$). Then we  substitute the found values of  $\tq^{EP}$ into (\ref{k0_except}) to obtain the corresponding exceptional points responsible for   collisions of the eigenvalues    (\textcolor{black}{upper} panel in  Fig.~\ref{fig:except}).

We emphasize that  for the most exceptional points  shown in Fig.~\ref{fig:except} the \emph{increase} of the gain-and-loss coefficient $q_0$ for the fixed $\nu$ leads to that a pair of initially complex eigenvalues collide on the real axis at $q_0=q_0^{EP}$ and splits into a pair of real ones as $q_0>q_0^{EP}$. This situation is opposite to what takes place in many conventional $\PT$-symmetric systems where the increase of the gain-and-loss       usually  destabilizes  the system, i.e., leads to the emergence of new complex eigenvalues (see \cite{Cart,BZ,Cartarius} and \cite{KYZ} for examples of this behavior). Only a few exceptional points  (labeled with crosses in Fig.~\ref{fig:except}) lead to the emergence of new complex eigenvalues as $q_0$ exceeds the exceptional point $q_0^{EP}$.

\section{Currents supported by spectral pseudo-singularities}
\label{sec:const_ampl}

{\color{black} Turning now   to the nonlinear case we focus on   (\ref{GPE}) with $\nu=0$, because  otherwise   the underlying linear system is in the broken $\PT$-symmetric phase, and its spectrum is partially complex. Furthermore, }
for the next consideration it is important to notice that there is  another essential difference between eigenvalues $\tq_0$ and $\tq_n$   in (\ref{spectrum_0}) and (\ref{spectrum_n}), besides the different multiplicity    mentioned above. The eigenfunction corresponding to $\tq_0$ is a plane wave with    {\em constant} amplitude, generating a constant flow from the lasing potential to absorbing one (schematically shown in Fig.~\ref{fig:one}), while the eigenfunctions corresponding to $\tq_n$ are modes with $\theta$-dependent amplitudes. This, in particular, means that in the nonlinear case, the modes having wavenumbers equal to $\tq_0$ will also have constant amplitudes, i.e., they will represent constant-amplitude nonlinear currents, with modified chemical potential $\mu$. Indeed, one can easily find an exact stationary (plane wave) solution  of the nonlinear equation \eqref{GPE} in the form
 \begin{eqnarray}
 \label{plane_wave}
 \Psi_1^{pw}(\theta,t)=\rho e^{-iq_0\theta-i\mu_0 t}, \quad  \Psi_2^{pw}(\theta,t)=\rho e^{iq_0\theta-i\mu_0 t}
 \end{eqnarray}
 where $\rho\geq 0$ is the arbitrary uniform amplitude, and  the chemical potential is given by ``nonlinear dispersion relation''
 \begin{eqnarray}
 	\label{mu0}
 \mu_0=\rho^2+q_0^2.
\end{eqnarray}
According to the notations used in Eqs.~(\ref{boundary}), subscripts 1 and 2 describe the solution in the intervals $[-\pi, 0)$ and $(0, \pi)$, respectively.

Thus, in spite of the dissipative character of the system the amplitude $\rho$ can be arbitrary.
Another feature of the obtained nonlinear currents (\ref{plane_wave}) is that the wavevector is defined by the (direct or time-reversal) pseudo-singularity $q_0$, rather than by the quantization condition imposed by the circular geometry (as it happens with all other eigenmodes).

\begin{figure}
	\begin{center}
		\includegraphics[width=0.8\columnwidth]{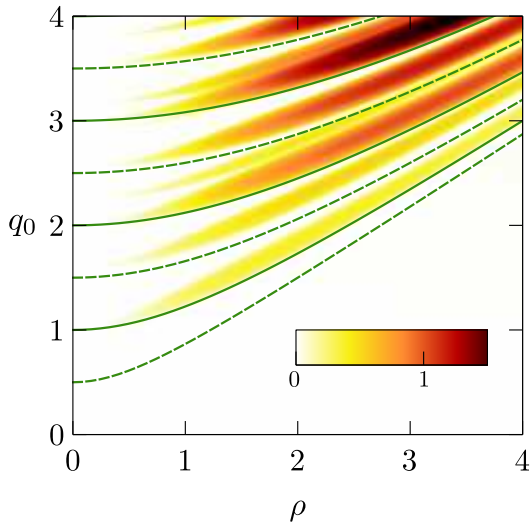}
	\end{center}
	\caption{(Color online) Instability increment $\max \{\textrm{Re}\lambda\}$ of the constant-intensity solutions  on the plane $\rho$ vs. $q_0$. Solid (dashed) lines correspond to Eq.~(\ref{eq:bifcond})  with $m=2,4,6$ ($m=1,3,5,7$), see  the  related discussion in Sec.~\ref{sec:bifurc}. }
	\label{fig:stab}
\end{figure}

\begin{figure}
 	\begin{center}
 		\includegraphics[width=0.99\columnwidth]{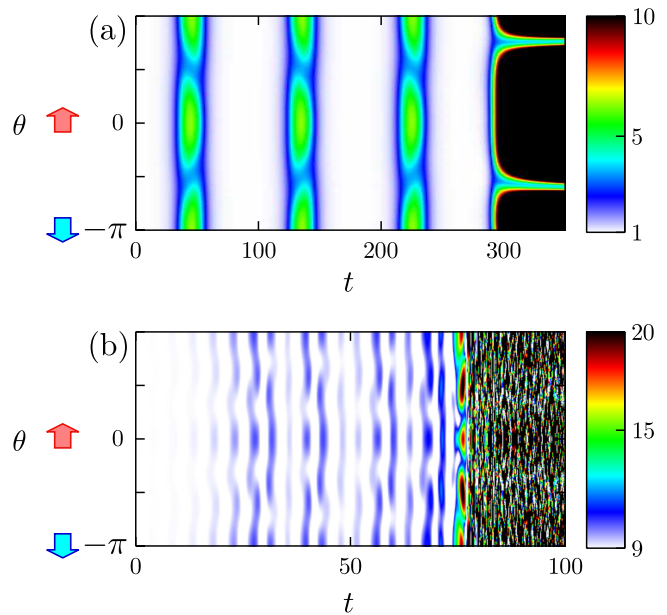}
 	\end{center}
 	\caption{(Color online) Evolution of two unstable plane waves: with $\rho=1$ and $q_0=1.4$ (a) and $\rho=3$ and $q_0=2.7$ (b). Shown are the pseudo-color plots of densities $|\Psi|^2$. Upward (downward) pointing arrows show  the position of the active (absorbing) potentials. }
 	\label{fig:dyn01}
 \end{figure}

Using the substitution $\Psi(\theta,t) = \Psi^{pw}(\theta, t) +  [u(\theta)e^{\lambda t}  + v(\theta)e^{\lambda^* t}]e^{-i\mu t}$, $|u,v|\ll 1$, we have performed numerically the standard linear stability analysis computing values of $\lambda$, i.e., the growth rates of the eventual instabilities.   The outcomes of this calculation are presented in the  stability diagram  in Fig.~\ref{fig:stab}.
\textcolor{black}{For sufficiently small densities $\rho$ all the currents are stable, which agrees with the fact that the spectrum of the corresponding linear problem is purely real, and hence all the linear modes (and  nonlinear modes of sufficiently small amplitude) are stable} (or, speaking more precisely, do not allow for   the exponential growth of the perturbation). For nonlinear currents with larger densities,   the stability diagram features    nontrivial striped structure.  We observe remarkable stability of backgrounds of relatively high amplitude $\rho$ for $q_0$ below $0.5$, i.e.,  below the lowest bifurcation point (see Sec.~\ref{sec:bifurc}). Surprisingly, the stability domain increases with the amplitude background, i.e. the nonlinearity appears to be a stabilizing factor. Moreover, for a given background amplitude the stability and instability domains alternate. 
The unstable eigenvalues can be either purely real [say, for the solution illustrated in Fig.~\ref{fig:dyn01}(a) unstable eigenvalue is $\lambda\approx0.20$] or form a complex-conjugate pair [for the solution   in Fig.~\ref{fig:dyn01}(b) unstable eigenvalues are $\lambda\approx0.20\pm 1.33i$].

The dynamics of unstable modes is illustrated in Fig.~\ref{fig:dyn01}. We observe that at early stages of the instability development   the initially uniform  density   becomes nearly periodic in space (and time). The spatial period is close to $\pi$ and is defined by the location of $\delta$  functions on the circle (separated by $\pi$).  For larger times the mean density     grows and for larger $q_0$ and $\rho$ [Fig.~\ref{fig:dyn01}(b)] becomes highly irregular for large times.

\section{Nonlinear periodic solutions}
\label{sec:periodic}

\subsection{General formulas}

As we observed in Fig.~\ref{fig:dyn01}, transient spatially periodic patterns emerge at intermediate  stages  of unstable evolution of nonlinear currents with initially uniform density.  This observation raises a natural question about the existence of truly stationary and dynamically stable  modes with periodic density in this system. In this section, we will show that such periodic modes indeed exist and, moreover, they can be found in the explicit analytical form. To this end, it is convenient to reformulate the problem such that it becomes invariant under $\PT$ transformation, where ${\cal P}$ is the parity reversal with respect to $\theta=0$: ${\cal P}\Psi(\theta,t) = \Psi(-\theta,t)$, and ${\cal T}$ is the time reversal ${\cal T}\Psi(\theta,t) = \Psi^*(\theta, -t)$. Therefore,   now we place    $\delta$-functions  at $\theta = \pm \pi/2$ and impose periodic boundary conditions at $\theta=\pm \pi$. (Obviously, the newly obtained system is equivalent to that considered in Sec.~\ref{sec:const_ampl}: one can be transformed to another by the shift of $\theta$ by $\pi/2$; hence all the properties remain invariant and the   difference is only in the     mathematical expression of the results). Thus for stationary modes $\Psi(\theta, t)= \psi(\theta) e^{-i\mu t}$ we have
\begin{equation}
\label{GPE_1}
-\psi_{\theta\theta} + |\psi|^2\psi + 2iq_0[\delta(\theta+\pi/2)-\delta(\theta- \pi/2)]\psi = \mu \psi.
\end{equation}

We will look for $\PT$-symmetric solutions, i.e.,
$\psi(\theta) = \psi^*(-\theta)$, and use the Madelung representation $\psi = \rho(\theta)e^{i\phi(\theta)}$. Here $\rho(\theta)$ is the amplitude of the solution; in the case $\rho(\theta)=\textrm{const}$, we recover the constant-amplitude solutions described in Sec.~\ref{sec:const_ampl}. Defining also the current density  $j=\phi_\theta \rho^2$, the equation for the amplitude of the field can be integrated, which gives
\begin{equation}
\label{eq:Madelung}
\rho_{\theta}^2 + {\mu} \rho^2 - \frac{1}{2}\rho^4 +\frac{j_0^2}{\rho^2}=\frac{E}{2}.
\end{equation}
where $E$ is an integration constant, and the current is  piecewise-constant:
\begin{eqnarray}
\label{current}
j(\theta)=\left\{
\begin{array}{ll}
j_0 &\mbox{ for $\theta \in (-\pi/2, \pi/2)$}
\\
-j_0 &\mbox{ for $\theta \in [-\pi,-\pi/2)\cup (\pi/2,\pi)$}
\end{array}
\right.
\end{eqnarray}

Further analysis is performed using the fact that the $\rho(\theta)$ is an even function, $\rho(\theta)=\rho(-\theta)$,
and can be searched in a form of squared Jacobi elliptic functions. Specifically, we explore two substitutions as follows
\begin{subequations}
\label{eq:dn}
\begin{eqnarray}
\label{eq:dn-}
\rho_-^2 &=&  \mu - C + B - 2B\dn^2(\sqrt{B}\theta, k),\\
\label{eq:dn+}
\rho_+^2 &=&  \mu - C + B - 2B\dn^2(K - \sqrt{B}\theta, k),
\end{eqnarray}
\end{subequations}
where $K=K(k)$ is the complete elliptic integral of the firsts kind, and $k$ stands for the elliptic modulus; $B$ and $C$ are positive  parameters which have to satisfy the relations
\begin{eqnarray}
\label{eq:k}
k^2 &=& (B + 3C-\mu)/(2B),
\\
\label{eq:j02}
j_0^2 &=&  {C(\mu - C - B)(\mu - C + B)}.
\end{eqnarray}
The integration constant is computed as
\begin{eqnarray}
\label{eq:E}
E = -B^2-3C^2+2C\mu+\mu^2.
\end{eqnarray}

The main difference between solutions $\rho^2_-$ and   $\rho^2_+$ is   readily seen: one of them ($\rho_-^2$) corresponds to   solutions with a local density minimum at the origin $\theta=0$, whereas solutions of another type ($\rho_+^2$) has a local density maximum at the   $\theta=0$.

Even though the problem is considered on the circle $\theta \in [-\pi, \pi)$, we require the density $\rho^2(\theta)$ to be $\pi$-periodic (not just $2\pi$-periodic):
\begin{eqnarray}
\label{cyclic}
 \rho^2(\theta) = \rho^2(\theta +\pi),
\end{eqnarray}
where $\rho$ stays for  $\rho_-$ or $\rho_+$, depending on the  particular choice of the solution in (\ref{eq:dn}).
On the one hand, $\pi$-periodic density   means that the half-rings $\theta\in[-\pi/2, \pi/2)$ and $\theta\in(-\pi,-\pi/2)\cup (\pi/2, \pi)$ have equal density distributions; on the other hand, this  guarantees automatically that the full complex-valued solution $\psi(\theta)$ is $2\pi$-periodic, provided that the current $j(\theta)$   satisfies (\ref{current}) [see Sec.~\ref{sec:bifurc} below for a comment on the existence of solutions with $2\pi$-periodic, but not $\pi$-periodic densities].

From the periodicity constraint (\ref{cyclic}) we   have the quantization condition $B=B_m$ where
\begin{equation}
\label{eq:1B}
B_m = K^2m^2/\pi^2, \quad m=2,4,6\ldots
\end{equation}
Then     $m$ counts the number of periods of density distribution.

Finally, in order for the solutions (\ref{eq:dn}) to respect the matching conditions   at $\theta=\pm\pi/2$, the following relation   between the current and field densities must hold:
\begin{eqnarray}
\label{eq:j0}
j_0=q_0\rho^2(\pi/2).
\end{eqnarray}

\subsection{Non-zero current solutions}

Let us now consider several particular cases for which auxiliary relations from the previous subsection   provide physically meaningful periodic solutions.

\subsubsection{Solutions with local maxima at the spots with the gain and losses}

In this subsection, we consider the case when the density $\rho^2(\theta)$ is given  by the expression for $\rho_-^2$ in (\ref{eq:dn-}) for $m=2, 6, 10, \ldots$ \textit{or} by the expression for $\rho_+^2$ in (\ref{eq:dn+}) for $m=4, 8, 12,\ldots$. In order to use (\ref{eq:j0}), we evaluate the densities at $\theta=\pi/2$ and observe that for the chosen values of $m$ the density is given for the same expression either for $\rho_+^2$ or for $\rho_-^2$: $\rho^2(\pi/2) = \mu-C-B-2B(1-k^2)$.  Then  the  combination of (\ref{eq:k}),(\ref{eq:j02}), and  (\ref{eq:j0})      gives
\begin{equation}
\label{eq:quad1}
(\mu-C-B)(\mu-C+B)=4q_0^2C.
\end{equation}
Next, one can use equations  (\ref{eq:k}) and (\ref{eq:1B})  to express $B$ and $C$ trough $k$, $K(k)$, and $\mu$, and  transform  (\ref{eq:quad1}) into a quadratic equation with respect to $\mu$, resulting in two branches of $\mu(k)$ parametrized by the elliptic modulus  $k\in [0, 1)$. Physically relevant solutions belong to one branch given as
\begin{eqnarray}
\label{eq:mumax}
\mu(k;m) &=& \frac{K^2m^2(2k^2-1)}{2\pi^2} + \frac{3q_0^2}{2} +
\nonumber \\
&+& \frac{3q_0^2}{2}\sqrt{\left(1- \frac{K^2m^2}{\pi^2q_0^2}\right)^2 + \frac{4k^2K^2m^2}{\pi^2q_0^2}}.
\end{eqnarray}
The second branch of solutions, with ``$-$'' sign in front of the radical in (\ref{eq:mumax}), corresponds to non-physical solutions with negative amplitudes.

The densities of the found solutions can be computed as follows. For $m=2,6, \ldots$
 the field intensity  reads
\begin{equation}
\label{eq:cn-}
\rho_-^2 =  \frac{2\mu}{3}  +\frac{2K^2m^2}{3\pi^2}\left[3k^2\sn^2\left(\frac{Km}{\pi}\theta,k\right) - k^2-1\right].
\end{equation}
while for $m=4,8, \ldots$ the density is
\begin{equation}
\label{eq:cn+}
\rho_+^2 =  \frac{2\mu}{3}   +\frac{2K^2m^2}{3\pi^2}\left[3k^2\sn^2\left(\!K - \frac{Km}{\pi}\theta,k\!\right) - k^2-1\right].
\end{equation}
For solutions of either type,    the current density reads
\begin{equation}
\label{eq:j0max}
j_0 = \frac{2q_0}{3}\left[\mu + \frac{K^2m^2}{\pi^2}(2k^2-1)\right].
\end{equation}

\begin{figure}
	\begin{center}
		\includegraphics[width=\columnwidth]{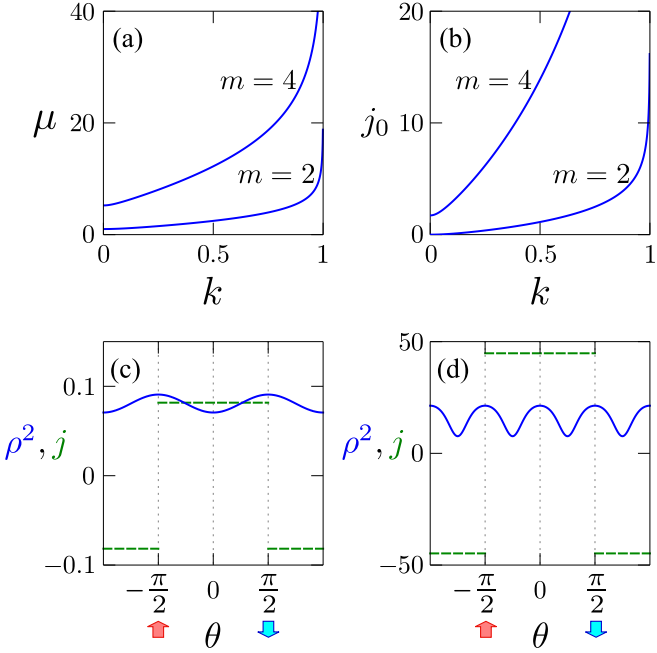}
	\end{center}
	\caption{(Color online) Chemical potential  (a)  $\mu(k)$  from (\ref{eq:mumax})  and current density (b) $j_0(k)$  from (\ref{eq:j0max}) for two families of solutions with maxima at the lossy and active spots corresponding to $m=2$ and $q_0=0.9$ and $m=4$ and $q_0=2.1$. Lower panels show examples of solutions: (c) corresponds to $m=2$ and $k=0.1$, and (d) corresponds to $m=4$ and $k=0.9$. Densities $\rho^2(\theta)$ and currents $j(\theta)$ are shown with solid and (horizontal) dashed lines, respectively. Upward (downward) pointing arrows show  the position of the active (absorbing) potentials.}
	\label{fig:mumax}
\end{figure}

The obtained solutions are illustrated in Fig.~\ref{fig:mumax}. All solutions of this type are characterized by local maxima of the density at $\theta=\pm \pi/2$, i.e., at the points where the gain and the dissipation are placed. At $\theta=0$ the behavior is different:  for the given $m$ the distribution $\rho_-^2$  ($\rho_+^2$) corresponds to solution with a local density minimum (maximum) at $\theta=0$.

For each $m$, the solutions exist for any $q_0\geq 0$ and $k\in [0, 1)$. As $k$ approaches 1, the chemical potential $\mu$ and the current  density $j_0$ diverge, see Fig.~\ref{fig:mumax}(a,b). In the opposite limit, $k=0$, the solution can feature different behavior: in some cases [see the curves with $m=2$ in Fig.~\ref{fig:mumax}(a,b)] the density and the current vanish at $k=0$. However, it can also happen that the current is finite and nonzero at $k=0$  [see curves with $m=4$ in Fig.~\ref{fig:mumax}]. The detailed understanding of the case $k=0$ is presented below, in Sec.~\ref{sec:bifurc}.

We have systematically scanned the domain $(q_0, k) \in [0,5]\times [0, 1)$ with $m=2$ and $m=4$ and observed that all the tested solutions are stable.

\subsubsection{Solutions with local minima at the spots with the gain and losses}

Now we choose the opposite combination of $m$ and $\rho_\pm$:  we assume that   the density $\rho^2(\theta)$ is given  by the expression for $\rho_-^2$ in (\ref{eq:dn-}) for $m=4, 8, 12, \ldots$  {or} by the expression for $\rho_+^2$ in (\ref{eq:dn+}) for $m=2, 6, 10, \ldots$. Then the density at $\pi/2$ can be evaluated as   $\rho^2(\pi/2) = \mu-C-B$, and the combination of   (\ref{eq:k}),  (\ref{eq:j02}), and  (\ref{eq:j0})      yields the relation
\begin{equation}
\label{eq:quad2}
C(\mu-C-B)(\mu-C+B)=q_0^2(\mu-C-B)^2.
\end{equation}
The particular case $\mu-C-B=0$ corresponds to the solutions with zero currents which are considered below in Sec.~\ref{sec:0cur}. For $\mu-C-B\ne0$, Eq. (\ref{eq:quad2}) again transforms into a quadratic equation for $\mu$ which gives two physically relevant branches of solutions:
\begin{eqnarray}
\label{eq:mumin}
%
\mu_\pm(k;m) = -\frac{K^2m^2(1+k^2)}{2\pi^2}  + \frac{3q_0^2}{2} \pm \hspace{1cm}
\nonumber \\
 \frac{3q_0^2}{2}\sqrt{\left( 1 - \frac{K^2m^2(k^2+1)}{\pi^2 q_0^2}\right)^2 - \frac{4K^4k^2m^4}{\pi^2q_0^2}}.
\end{eqnarray}
The densities $\rho_-^2(\theta)$ for    $m=4, 8, 12, \ldots$ and   $\rho_+^2$   for $m=2, 6, 10, \ldots$ can be again found from (\ref{eq:cn-}) and (\ref{eq:cn+}), respectively.

The solutions obtained in  this way are characterized by the local  density minima at the points of the gain and dissipation $\theta=\pm\pi/2$. As follows from (\ref{eq:mumin}), for each $m$ there exist two families of such solutions with different chemical potentials. The densities of the two families with the chemical potential (\ref{eq:mumin}) are simply shifted on a certain constant with respect to each other. Hence {the respective solutions can be conveniently termed large-density modes ($\mu_+$) and small-density modes ($\mu_-$)}. It is remarkable  that for such solutions to exist, the gain-and-loss strength $q_0$ must be sufficiently large, namely,
\begin{equation}
\label{eq:qth}
q_0 \geq q_{th} = \frac{Km}{\pi}(k+1).
\end{equation}
At $q_0 = q_{th}$ the square root in (\ref{eq:mumin}) vanishes, at this instant $\mu_+=\mu_-$, and the two families merge.

The   currents are given by
\begin{equation}
j_0^\pm = \frac{2q_0}{3}\left[\mu_{\pm}(k; m) - \frac{K^2m^2}{\pi^2}(k^2+1)\right].
\end{equation}

\begin{figure}
	\begin{center}
		\includegraphics[width=\columnwidth]{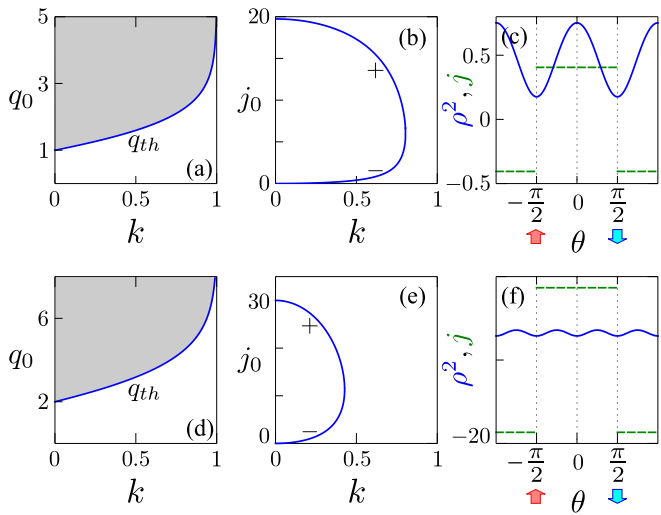}
	\end{center}
	\caption{(Color online) Solutions with local minima at the spots with  the gain and losses. Panels (a)--(c) illustrate the case $m=2$:  domain of existence shaded   (a),   dependencies $j_0(k)$ for $q_0=2.3$ (b), profile   {of the small-density solution} with $k=0.5$. Panels (d)--(f) illustrate the case $m=4$ and are organized in the same way,  {but now the dependencies in panel (e) correspond to $q_0=3$, and in (f) the large-density solution with} $k=0.4$ is shown. Labels ``$+$'' and ``$-$'' in (b) and (e) mark families of large-density (``$+$'') and small-density (``$-$'') solutions. Upward (downward) pointing arrows show  the position of the active (absorbing) potentials.}
	\label{fig:mumin}
\end{figure}

Solutions with local minima at the points of gain and losses are illustrated in Fig.~\ref{fig:mumin}. We have checked solutions with $m=2$ and $m=4$ and observed that all solutions of this type are stable in the whole domain of their existence.

\subsection{Zero-current solutions}
\label{sec:0cur}
Taking into account that the system at hand has   localized gain and dissipation spatially separated from each other, the existence of the modes carrying currents between the potentials appears to be natural. It turns out to be much less intuitive that the system may have  zero-current  solutions. Nevertheless, such solutions can be found: assuming $\mu=C+B$ in (\ref{eq:quad2}) and performing simple transformations, we eventually obtain the expressions for  real-valued wavefunctions in the following form:
\begin{subequations}
	\label{eq:sn}
\begin{equation}
\label{eq:sn1}
\psi_-(\theta) =  \frac{\sqrt{2}Kmk}{\pi}\sn\left(\frac{Km\theta}{\pi}, k\right), \quad m=4, 8, \ldots
\end{equation}
\begin{equation}
\label{eq:sn2}
\psi_+(\theta) =  \frac{\sqrt{2}Kmk}{\pi}\sn\left(K-\frac{Km\theta}{\pi}, k\right), \quad m=2, 6, \ldots
\end{equation}
\end{subequations}
Both types of the solutions in (\ref{eq:sn}) correspond to the chemical potential
\begin{equation}
\mu  = \frac{K^2m^2}{\pi^2}(1+k^2).
\end{equation}

\begin{figure}
	\begin{center}
 \includegraphics[width=0.99\columnwidth]{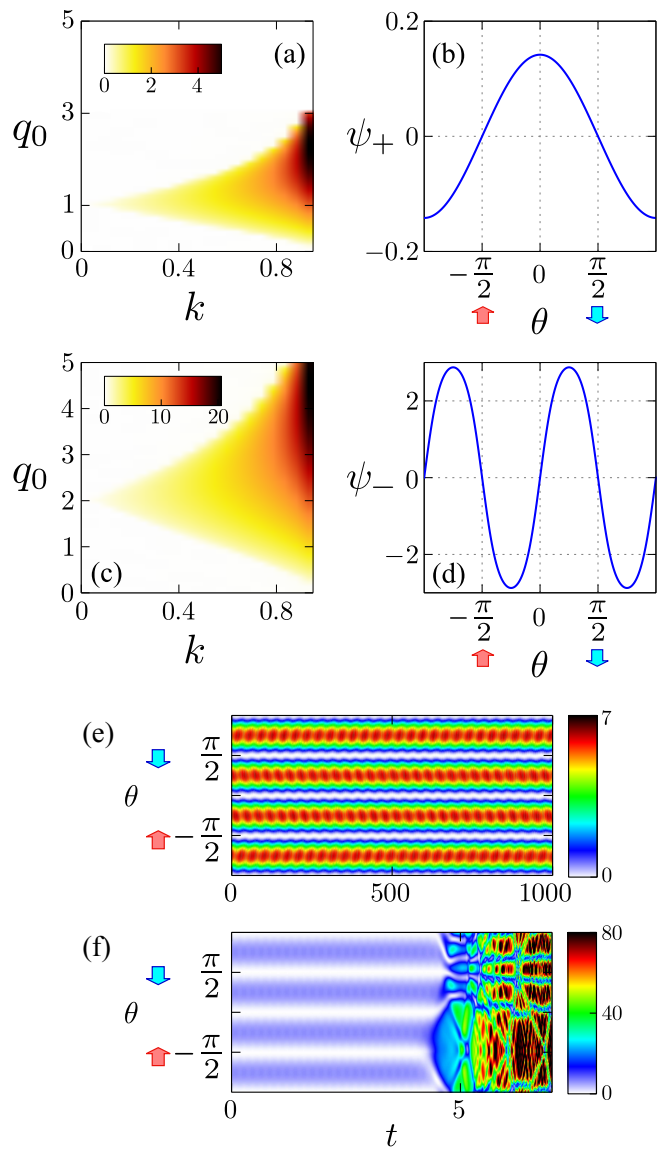}
 	\end{center}
 	\caption{(Color online) Solutions with zero current. Panels (a) and (b) correspond to $\psi_+$-solutions (\ref{eq:sn2}) with $m=2$: pseudo-color plot of the instability increment  $\max \{\textrm{Re}\lambda\}$  (stable solutions correspond to the white domain), and example of a solution with $k=0.1$ (note that the shape of solutions does not depend on $q_0$). Panels (c) and (d) correspond to  $\psi_-$-solutions (\ref{eq:sn1}) with $m=4$. The solution shown in (d) has $k=0.8$. Notice that the   solutions are exactly zero at $\theta=\pm\pi/2$. Panels (e) and (f) show stable (e, $q_0=5$) and unstable (f, $q_0=3$) evolution of the mode  from (d). Upward (downward) pointing arrows show  the position of the active (absorbing) potentials.}
 	\label{fig:0curr}
 \end{figure}

Physically, understanding of the existence of the  currentless solutions stems from the fact that their density is \emph{exactly} zero at $\theta=\pm \pi/2$, and therefore either gain or loss potentials, and hence the value of $q_0$, do not affect these   solutions (this is a situation similar to one known for the dark solitons~\cite{BKPO} in dissipative waveguides). Obviously, the current carried  by such modes is identically zero irrespectively of the  gain-and-loss strength $q_0$. However, the value of $q_0$ affects the stability of solutions. This is illustrated by stability  diagrams in Fig.~\ref{fig:0curr}. The solutions are generically stable for small elliptic moduli $k$ (i.e., for small-amplitude densities) but can lose stability as $k$ increases (i.e., the amplitude of  $|\psi|^2$ grows). For $q_0=m/2$ these solutions are unstable for any $k$.

The  existence of stable solutions with zero current can look  surprising, because such solutions require the nodes of the wavefunction to coincide exactly with the points where the gain and dissipation are situated. However, the remarkable robustness of such solutions has been also confirmed by simulations of their evolution. We prepare the initial condition in the form of a stationary mode $\psi(\theta)$ slightly shifted in $\theta$ [such that its nodes no longer coincide exactly with $\pm\pi/2$] and run the evolution. The stable mode perturbed in this way  ``breathes'' slightly, but   preserves its distinctive shape for indefinitely long evolution, see Fig.~\ref{fig:0curr}(e). Reducing $q_0$ from 5 to 3, we make the same mode unstable [see the stability diagram in Fig.~\ref{fig:0curr}(c)]; in this case the mode destroys rapidly: first, the instability manifests itself in the growth of the intensity in the region adjacent to the active spot, that is around $\theta=-\pi/2$. Eventually the amplitude becomes large and highly irregular.


\subsection{Bifurcations from the plane waves}
\label{sec:bifurc}


All the solutions with periodic density described above  can be considered in the limit of small elliptic modulus, $k\to 0$. In this limit, the solution densities   become spatially uniform (i.e., $\theta$-independent). Analysis of this limit allows for clarification of the mutual relations among all the solutions. To consider the mentioned relations, let us start with solutions (\ref{eq:mumax}) having local maxima at the spots with dissipation and gain. Substituting $k=0$ in the respective equations, we see that  if $q_0\leq m/2$, then the chemical potential (\ref{eq:mumax}) in the limit $k=0$ amounts to $\mu(0; m)=m^2/4$, and the amplitude of the solution vanishes $\rho(\theta)\equiv 0$. In other words, the solution becomes a plane wave of infinitesimal amplitude $\rho=0$.
For $q_0\geq m/2$,  the chemical potential becomes
\begin{equation}
\mu(0;m)=3q_0^2 - \frac{m^2}{2},
\end{equation}
and the uniform density $\rho$ is nonzero at $k=0$. Using the expression (\ref{mu0}) for the chemical potential of modes with   constant densities, we find a set of ``quantized'' amplitudes $\rho_m$:
\begin{equation}
\label{eq:bifcond}
\rho_m^2=2q_0^2 - \frac{m^2}{2}.
\end{equation}
In other words, expression (\ref{eq:bifcond}) describes the solutions with constant densities, from which solutions with periodic and nonzero  amplitude bifurcate.

\begin{figure}[t]
	\begin{center}
 \includegraphics[width=0.8\columnwidth]{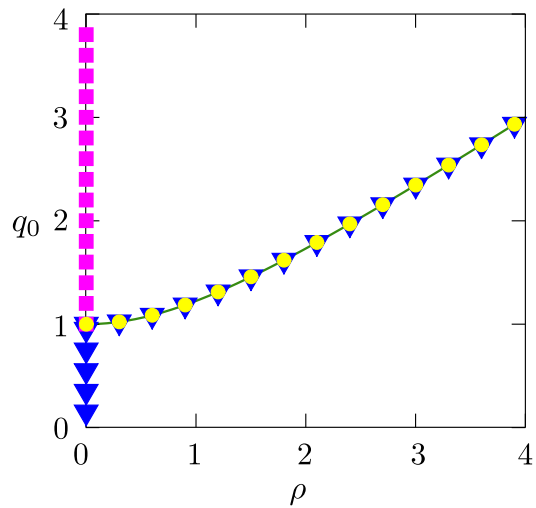}%
 	\end{center}
 	\caption{(Color online) Schematic bifurcation diagram in the limit $k=0$. Only solutions with $m=2$ are illustrated. Different labels  show the points in the plane $(q_0, \rho)$ from which  different solutions bifurcate. Blue  triangles correspond to  solutions with  local maxima at the points with gain and losses. Yellow circles  (magenta squares) correspond the large-density (small-density) solutions with    local  minima at the points with   gain and losses. The curve segment with overlapping triangles and circles corresponds to the points at which solutions of two different kinds bifurcate.}
 	\label{fig:bif}
 \end{figure}

The obtained results are illustrated in  Fig.~\ref{fig:bif}, where blue triangles show the locus of the points in the diagram $(\rho, q_0)$ where bifurcations of periodic solutions (with $m=2$) from the plane waves take place: one segment of this curve has $\rho=0$ ($q_0\leq m/2$), and another segment has $\rho>0$ ($q_0>m/2$).

For solutions with minima at the points of dissipation the situation is different. As follows from (\ref{eq:qth}), in the limit $k=0$ such solutions exist only if $q_0\geq m/2$. Substituting $k=0$ in Eq.~(\ref{eq:mumin}), we observe that the large-density solutions   [``$+$'' sign in (\ref{eq:mumin})] bifurcate from nonzero plane waves at the  points given by  (\ref{eq:bifcond}): hence yellow circles in Fig.~\ref{fig:bif} overlap with blue triangles. The small-density solutions with small densities [``$-$''  in (\ref{eq:mumin})] bifurcate from the infinitesimal-amplitude plane waves (magenta squares in Fig.~\ref{fig:bif}).

Finally, the zero-current solutions (\ref{eq:sn})  in the limit $k=0$ transform into  the  plane waves of infinitesimal amplitude for any $m$ and $q_0$.

It is interesting that in the diagram in Fig.~\ref{fig:stab} the domains of instability of the plane waves are  ``separated'' by bifurcation condition (\ref{eq:bifcond}).
Note that in Fig.~\ref{fig:stab}, apart from   the curves (\ref{eq:bifcond})  with even  $m$  (solid lines), we have also plotted  curves (\ref{eq:bifcond})  with odd $m$ (dashed lines). While the exact solutions presented above   correspond only to even $m$, the system might,  hypothetically speaking,  admit  solutions with odd $m$ as well. Such solutions, should they exist, would be  essentially different from the solutions with even $m$: for odd $m$  the density distribution at   half ring $\theta\in(-\pi/2, \pi/2)$ is not  identical to that on another half-ring. In other words, the  density $\rho^2(\theta)$ of such solutions  would be $2\pi/m$-periodic, but not $\pi$-periodic. As a result,  for solutions  with odd $m$ to be $2\pi$-periodic, an \emph{additional} condition appears which requires   the argument of the complex-valued function $\psi(\theta)$ to  match  periodically at $\theta=-\pi$ and $\theta=\pi$. This condition (which can be expressed analytically in terms of the elliptic integral of the third kind) is satisfied automatically for solutions with even $m$, but, generically speaking, is not satisfied for odd $m$. This additional condition makes solutions with odd $m$ ``difficult to find'': so far we have not been able to find any solution with odd $m$ (in a sense this situation reminds of  the issue of the existence of asymmetric localized modes in $\PT$-symmetric systems which are known to exist only under a very particular choice of the $\PT$-symmetric potential \cite{KYZ}).

 \section{Conclusion}

To conclude, we described in details a ring-shaped waveguide with spatially localized gain and losses.  The localized complex potentials were modeled by the Dirac $\delta$ functions with imaginary amplitudes. An imaginary $\delta$ function considered on the infinite line has a spectral singularity in its continuum spectrum. A configuration of two delta functions on the ring does not have a spectral singularity in its strict mathematical sense (at least because the spectrum of the system on the ring is entirely discrete). However, bearing in mind the analogy between the properties of the spectral singularities and the ``lasing'' and ``absorption'' in our system, we say that our configuration has a pseudo-(spectral)-singularity characterized by the imaginary amplitude of the $\delta$ functions.

We have started our analysis from the linear case computing the spectrum of the linear waves.  If the $\delta$ functions are places symmetrically on the ring, then the linear spectrum is always real. However, the deviation of the symmetric placement of the gain and loss enables the possibility of the existence of the exceptional points. Extending the consideration onto the nonlinear setup,  we have found a variety  of nonlinear modes  many of which are dynamically stable. A particularly interesting class of solutions corresponds to the plane   (constant-amplitude) waves which feature several relevant properties. First, their wavevectors correspond to the wavelength of the spectral singularities of the $\delta$ function potentials. Second, the stability diagram of these plane waves features the stripe-patterned structure with alternating domains of stable and unstable solutions. Third, the plane waves     can originate solutions with spatially periodic densities. The latter have been found  in the explicit analytical form and classified depending on their shapes. Another interesting peculiarity of our configuration is the existence of the periodic currentless solutions whose density vanishes at the spots with gain and dissipation. These solutions can be stable or unstable, depending on the value of the gain-and-loss strength.

 {\color{black} Finally, we make a remark on the role of $\PT$ symmetry in the reported system. For the existence of the stationary constant-amplitude patterns it was of crucial importance that laser and absorber   operated at the same wavelength. This was easily achieved using two delta function potentials which   differ   by the sign of their imaginary part. In the meantime,  one may find other shapes of the lasing and absorbing potentials, having distinct landscapes of the complex parts but respective spectral singularity and time reversed spectral singularity, at the same wavelength. Yet another possibility is to consider laser and absorber at different wavelengths, but embedded in a ring-shaped waveguide with varying dielectric constant. Clearly, both mentioned geometries do not obey $\PT$ symmetry and at the same time ensure emission and complete absorption of the radiation at different spatial locations. However, some of the properties of such a non-$\PT$-symmetric system can be sufficiently different (and worthy of a separate study) from those of our $\PT$-symmetric configuration. One of important  differences can be related to the absence of the  continuous families of solutions in the non-$\PT$-symmetric case. The stability of the existing solutions will be also strongly affected by the asymmetric gain-and-loss distribution. One  should expect that stable modes will eventually become isolated attractors, which is not the case of our system.}

\begin{acknowledgments}
	The work was supported by the FCT (Portugal) grant UID/FIS/00618/2013.
\end{acknowledgments}


\begin{thebibliography}{99}

\bibitem{Oppo1} R. Livi, R. Franzosi, and G.-L. Oppo, ``Self-Localization of Bose-Einstein Condensates in Optical Lattices via Boundary Dissipation,'' Phys. Rev. Lett. {\bf 97}, 060401 (2006).

\bibitem{BKPO} V. A. Brazhnyi, V. V. Konotop, V. M. P\'erez-Garc\'ia, and H. Ott, ``Dissipation-Induced Coherent Structures in Bose-Einstein Condensates,'' Phys. Rev. Lett. {\bf 102}, 144101 (2009).

\bibitem{Oppo2}  R. Campbell and G. L. Oppo,	
``Stationary and traveling solitons via local dissipations in Bose-Einstein condensates in ring optical lattices,''
arXiv:1606.06773 [nlin.PS]

\bibitem{ZK}  D. A. Zezyulin, V. V. Konotop, G. Barontini, and H. Ott,
``Macroscopic Zeno effect and stationary flows in nonlinear waveguides with localized dissipation,''
Phys. Rev. Lett. {\bf 109} 020405 (2012); D. A. Zezyulin and V. V. Konotop,
``Stationary vortex flows and macroscopic Zeno effect in Bose-Einstein condensates with localized dissipation,'' Rom. Rep. Phys. {\bf 67}, 223 (2015).

\bibitem{Ott} G. Barontini, R. Labouvie, F. Stubenrauch, A. Vogler, V. Guarrera, and H. Ott,
 ``Controlling the Dynamics of an Open Many-Body Quantum System with Localized Dissipation,''
 Phys. Rev. Lett. {\bf 110}, 035302 (2013).

\bibitem{BarZezKon}  I. V. Barashenkov, D. A. Zezyulin, and V. V. Konotop,
``Jamming anomaly in $\PT$-symmetric systems,''
New J. Phys.  {\bf 18}, 075015 (2016).

\bibitem{Mostafa_review} M. A. Naimark, ``Investigation of the spectrum and the expansion in eigenfunctions of a nonselfadjoint operator of the second order on a semi-axis,'' Tr. Mosk. Mat. Obs., {\bf 3}, 181 (1954);
A. Mostafazadeh and H Mehri-Dehnavi,
``Spectral singularities, biorthonormal systems and a two-parameter family of complex point interactions,''
J. Phys. A, {\bf 42}, 125303 (2009);
A. Mostafazadeh, ``Physics of Spectral Singularities,'' Geometric Methods in Physics, Trends in Mathematics 145-165 (Springer
International Publishing Switzerland, 2015) DOI: \verb"10.1007/978-3-319-18212-4_10".

\bibitem{Zero_resonance} A.  Mostafazadeh, ``Spectral singularities of complex scattering potentials and infinite reflection and transmission coefficients at real energies,'' Phys. Rev. Lett. {\bf 102}, 220402 (2009); Z. Ahmed, ``Zero width resonance (spectral singularity) in a complex PT-symmetric potential,'' J. Phys. A  {\bf 42}, 472005 (2009).

\bibitem{CPA} Y. D. Chong,  L. Ge, H. Cao, and A. D. Stone, ``Coherent perfect absorbers: Time-reversed lasers,'' Phys. Rev. Lett.  {\bf 105}, 053901 (2010); W. Wan, Y. Chong, L. Ge, H. Noh, A. D. Stone,  and H. Cao, ``Time-reversed lasing and interferometric control of absorption,'' Science  {\bf 331}, 889 (2011).

\bibitem{Bender} C. M. Bender and S. Boettcher,
``Real spectra in non-Hermitian Hamiltonians having $\PT$ symmetry,''
Phys. Rev. Lett. {\bf 80}, 5243 (1998);
C. M. Bender,
``Introduction to $\PT$-symmetric quantum theory,''
Contemp. Phys. {\bf 46},  277 (2005);
C. M. Bender, ``Making sense of non-Hermitian Hamiltonians,''
Rep. Prog. Phys. {\bf 70}, 947 (2007).

\bibitem{KYZ} V. V. Konotop, J. Yang, and  D. A. Zezyulin,
	``Nonlinear waves in $\PT$-symmetric systems,''
Rev. Mod. Phys.  {\bf 88},  35002 (2016).


\bibitem{Mostafa_self_dual} A. Mostafazadeh,
``Self-dual spectral singularities and coherent perfect absorbing lasers without $\mathcal {P}\mathcal {T}$-symmetry,''
J. Phys. A: Math. Theor. {\bf 45}, 444024 (2012).

\bibitem{CPA-laser} S. Longhi, ``$\PT$-symmetric laser absorber,'' Phys. Rev. A {\bf 82}, 031801 (2010); Y. D. Chong,  L. Ge, and A. D. Stone,  ``$\PT$-symmetry breaking and laser-absorber modes in optical scattering systems,'' Phys. Rev. Lett. {\bf 106}, 093902 (2011).


\bibitem{Mostafa_nonlinear} A. Mostafazadeh,
``Nonlinear spectral singularities for confined nonlinearities,'' Phys. Rev. Lett.  {\bf 110}, 260402 (2013);
A. Mostafazadeh,  ``Spectral singularities and CPA-laser action in a weakly nonlinear $\PT$-symmetric bilayer slab,'' Stud.  App. Math. {\bf 133}, 353 (2014).

\bibitem{toroid} C. Ryu, M. F. Andersen, P. Clade, V. Natarajan, K. Helmerson,
and W. D. Phillips, ``Observation of persistent flow of a Bose-Einstein condensate in a toroidal trap,'' Phys. Rev. Lett. {\bf 99}, 260401 (2007); C. N. Weiler, T. W. Neely, D. R. Scherer, A. S. Bradley, M. J. Davis, and B. P. Anderson, ``Spontaneous vortices in the formation of Bose–Einstein condensates,'' Nature (London) {\bf 455}, 948 (2008); K. Henderson, C. Ryu, C. MacCormick and M. G. Boshier, ``Experimental demonstration of painting arbitrary and dynamic potentials for Bose-Einstein condensates,'' New J. Phys. {\bf 11}, 043030 (2009);
A. Ramanathan, K. C. Wright, S. R. Muniz, M. Zelan, W. T. Hill III, C. J. Lobb, K. Helmerson,
W. D. Phillips, and G. K. Campbell, ``Superflow in a toroidal Bose-Einstein condensate: An atom circuit with a tunable weak link,'' Phys. Rev. Lett. {\bf 106}, 130401 (2011).

\bibitem{atom_laser} R. J. C. Spreeuw, T. Pfau, U. Janicke, and M. Wilkens, ``Laser-like scheme for atomic-matter waves,'' EPL {\bf 32}, 469 (1995); N. P. Robins, P. A. Altin, J. E. Debs, J. D. Close, ``Atom lasers: Production, properties and prospects for precision inertial measurement,'' Phys. Rep. {\bf 529}, 265 (2013).


\bibitem{ring_shaped_polarit} V. K. Kalevich, M. M. Afanasiev, V. A. Lukoshkin, K. V. Kavokin, S. I. Tsintzos, P. G. Savvidis, and A. V. Kavokin, ``Ring-shaped polariton lasing in pillar microcavities,'' J. Appl. Phys. {\bf 115}, 094304 (2014).

\bibitem{polariton} T.  Gao, P.S. Eldridge, T. C. H. Liew, S. I. Tsintzos, G. Stavrinidis, G. Deligeorgis, Z. Hatzopoulos, and P. G.  Savvidis,
``Polariton condensate transistor switch,''
{Phys. Rev. B} {\bf 85} 235102, (2012).

\bibitem{lasers} L. Feng, Z. J. Wong, R. Ma,
Y. Wang, and X. Zhang, ``Single-mode laser by parity-time
symmetry breaking,'' Science {\bf 346}, 972 (2014).
%

\bibitem{BludKon} Yu. V. Bludov and V. V. Konotop, ``Acceleration and localization of matter in a ring trap,'' Phys. Rev. A {\bf 75}, 053614 (2007).



\bibitem{Cart} H. Cartarius and G. Wunner, ``Model of a $\PT$-symmetric Bose-Einstein condensate in a $\delta$-function double-well potential,'' Phys. Rev.  A {\bf 86}, 013612 (2012); A.  L\"ohle, H.  Cartarius, D.  Haag, D.  Dast, J. Main, G.  Wunner,
``Stability of Bose-Einstein condensates in a PT-symmetric double-$\delta$ potential close to branch points,'' Acta Polytechnica {\bf 54}, 
    133 
    (2014).

 \bibitem{BZ}  I. V. Barashenkov and D. A. Zezyulin,   ``Nonlinear modes in the ${\cal PT}$-symmetric
	double-delta well  Gross-Pitaevskii equation,''
	Proceedings of the 15 Conference on Pseudo-Hermitian Hamiltonians in Quantum Physics, May 18-23 2015, Palermo, Italy (Springer Proceedings in Physics, 2016), pp. 123-142. DOI: \verb"10.1007/978-3-319-31356-6_8"

\bibitem{Cartarius} W. D. Heiss, H. Cartarius, G. Wunner, and J. Main, ``Spectral singularities in $\PT$-symmetric Bose--Einstein condensates'', J. Phys. A: Math. Theor. {\bf 46}, 275307 (2013).
    
    
\bibitem{Mostafa_delta} A. Mostafazadeh,
``Delta-function potential with a complex coupling,'' J. Phys. A: Math. Theor.  {\bf 39}, 13495 (2006).




\end{thebibliography}
\end{document}